\documentstyle[amsfonts,prl,aps]{revtex}
\begin{document}
%\narrowtext

\title{Still Baking}
\author{Nathan Salwen}
\address{Lyman Laboratory of Physics\\
Harvard University\\
salwen@fas.harvard.edu}
 \author{Ron Rubin}
\address{Mathematics Department\\
Massachusetts Institute of Technology\\
rubin@math.mit.edu}

\date{\today}

\maketitle

\begin{abstract}
We present here a simple proof of the non-existence of a non-periodic
invariant point for the quantum baker's map propagator presented in
\cite{RS}%
, for Planck's constant $h=1/N$ and $N$ a positive integer.
\end{abstract}

\pacs{Quantum Chaos}

\paragraph*{Introduction}

A periodic quantum propagator was recently proposed in ref. \cite
{RS} which solved the parity violating problem of the original quantization
due to Balazs and Voros (ref. \cite{BV}). We present here a simple proof of
the non-existence of a non-periodic invariant point for this quantum baker's
map, for Planck's constant $h=1/N$ and $N$ any positive integer.

We review briefly the quantization. The quantum algebra of observables is
restricted to the set bounded operators on $L^{2}\left( \mathbb{R}\right) $
generated by the quantization of the classical generators of functions on
a torus: $U=\exp \left( 2\pi i\widehat{x}\right) $ and $V=\exp \left( 2\pi
i\widehat{p}\right) $so that $UV=e^{4\pi ^{2}i\hbar }VU$. A quantum
propagator was constructed by quantizing the dynamics of a covering map on
the plane.The quantum dynamics induced on the algebra of observables for the
quantum torus is the quantum baker's map. The covering dynamics is given by
the following mapping of ${\mathbb{R}}^{2}\rightarrow {\mathbb{R}}^{2}$:
\[
\left( x^{\prime },p^{\prime }\right)
\left\{ 
\begin{array}{ll}
(2x,p/2), & (x,p)\in l\cap e_{p}; \\ 
(2x-1,p/2+1/2), & (x,p)\in r\cap e_{p}; \\ 
(2x+1,p/2+1/2), & (x,p)\in l\cap o_{p}; \\ 
(2x,p/2), & (x,p)\in r\cap o_{p},
\end{array}
\right. 
\]
where 
\begin{eqnarray*}
l &:&=\left\langle [0,1/2)+\mathbb{Z}\right\rangle \times \mathbb{R}, \\
r &:&=\left\langle [1/2,1)+\mathbb{Z}\right\rangle \times \mathbb{R}, \\
e_{p} &:&=\mathbb{R}\mathbf{\times }\left\langle [0,1)+2\mathbb{Z}%
\right\rangle , \\
o_{p} &:&=\mathbb{R}\mathbf{\times }\left\langle [1,2)+2\mathbb{Z}%
\right\rangle .
\end{eqnarray*}
A corresponding quantum propagator which returns the classical baker
covering dynamics as $\hbar \rightarrow 0$ was found:
\begin{equation}
F=S(L+e^{-i\widehat{x}/\hbar }R)(E_{p}+e^{-i\widehat{p}/2\hbar }O_{p}),
\label{propagator}
\end{equation}
where $S$ is the unitary stretching and shrinking operator $S^{\dagger }%
\widehat{x}S=2\widehat{x}$, and $S^{\dagger }\widehat{p}S=\widehat{p}/2$,
and 
\begin{eqnarray*}
L &:&=\int_{[0,1/2)+\mathbb{Z}}\left| x\right\rangle \left\langle x\right|
\;dx,\quad  \\
R &:&=\int_{[1/2,1)+\mathbb{Z}}\left| x\right\rangle \left\langle x\right|
\;dx, \\
E_{p} &:&=\int_{[0,1)+2\mathbb{Z}}\left| p\right\rangle \left\langle
p\right| \;dp, \\
O_{p} &:&=\int_{[1,2)+2\mathbb{Z}}\left| p\right\rangle \left\langle
p\right| \;dp.
\end{eqnarray*}

For $h=1/N$ the algebra generated by $U$ and $V$ has a natural center
generated by 
\[
X=U^{N}=e^{i\widehat{x}/\hbar },\quad Y=V^{N}=e^{i\widehat{p}/\hbar }.
\]
That is 
\[
\left[ X,Y\right] =\left[ X,U\right] =\left[ X,V\right] =\left[ Y,U\right]
=\left[ Y,V\right] =0.
\]
In ref. \cite{LRS} and \cite{KLMR}, this insight was used to show that $%
L^{2}\left( \mathbb{R}\right) $ can be decomposed via the following
eigenvalue problem:
\begin{eqnarray*}
X\Phi  &=&e^{2\pi i\theta _{1}}\Phi , \\
Y\Phi  &=&e^{2\pi i\theta _{2}}\Phi ,
\end{eqnarray*}
where $\theta =\left( \theta _{1},\theta _{2}\right) \in
{\mathbb{T}}^{2}$. The simultaneous subspace ${\mathcal{H}}_{\hbar
  }\left( \theta \right) $ of $X$ and $Y$ has dimension $N=1/h$, and
the following inner product over the fundamental domain
$D=[0,1]\subset {\mathbb{R}}$ was derived:
\begin{eqnarray}
&&\left( \Psi _{1}(\theta ),\Psi _{2}(\theta )\right) _{P}  \label{ip} \\
&=&\int_{0}^{1}\overline{\Psi _{1}(x,\theta )}(K\Psi _{2})(x,\theta )dx,
\end{eqnarray}
where
\begin{eqnarray*}
K\Psi _{2}(x,\theta ) &=&\int_{-\infty }^{\infty }K\left( x,y\right) \Psi
_{2}(y,\theta )dy, \\
K\left( x,y\right)  &=&\frac{\sin \pi N\left( x-y\right) }{\pi \left(
x-y\right) }e^{-\frac{\pi N}{2}\left( \left( x-y\right) ^{2}+i\left(
x-y\right) \right) }
\end{eqnarray*}
A normalized ``position state basis'' was found to be the set $\left\{ \Phi
_{m}^{(\theta )}\right\} $, with $0\leq m<N-1$ and
\begin{equation}
\Phi _{m}^{(\theta )}=\frac{e^{2\pi i\theta _{2}m/N}}{N^{1/2}}\sum_{k\in %
\mathbb{Z}}e^{2\pi i\theta _{2}k}\left| \frac{\theta _{1}+m}{N}%
+k\right\rangle _{x}.  \label{delta comb}
\end{equation}
These are the $\delta $-comb wavefunctions seen for example in ref.
\cite{HB}%
.

The point $\theta =\left( 0,0\right) $ of the ``$\theta $-torus''
corresponds to the $N$-dimensional vector space ${\mathcal{H}}_{\hbar }\left(
0\right) $ of periodic $\delta $-combs. For the quantum baker's map, it was
shown that $\theta =$ $\left( 0,0\right) $ is an invariant point of the
dynamics on the $\theta $-torus \textit{for }$N$\textit{\ even}. That is,
the set of periodic $\delta $-combs is mapped onto itself by $F$. Restricted
to this subspace, the propagator $F$ is given by the following matrix
operator
\begin{equation}
\left( {\mathcal{Z}}\right) \left( {\mathcal{F}}^{N}\right)
{1}\left( 
\begin{array}{ll}
{\mathcal{F}}^{N/2} & 0 \\ 
0 & -{\mathcal{F}}^{N/2}
\end{array}
\right) \left( {\mathcal{Z}}^{-2}\right),   \label{OURS}
\end{equation}
where ${\mathcal{F}}_{nm}^{N}$ $=N^{-1/2}\exp \left( 2\pi inm/N\right) $
is the $N\times N$ discrete Fourier transform matrix
${\mathcal{Z}}_{nm}=\delta _{nm}\exp({i\pi n/N}).$ The form of
${\mathcal{Z}}_{nm}$ may be misleading for $n$ and $m$ outside of the
fundamental range $\left[ 0,N-1\right] $. In general, we write $\left(
  {\mathcal{Z}}\right) _{nm}=\delta _{nm}\exp({i\pi \left( n/N-\left[
      n/N\right] \right) })$, where $\left[ n/N\right] $ represents the
integer part of $n/N$.

\paragraph*{Are there more invariant points?}

An antiperiodic quantization scheme proposed by Saraceno in ref. \cite{S}
leads to the question of whether other points on the $\theta $-torus (for
instance $\theta =\left( 1/2,1/2\right) $ corresponding to anti-periodic
boundary conditions) are invariant. We demonstrate here that for the
propagator \ref{propagator} this will never occur. 

Assume there is such an invariant point. Then by definition, there exists
some value of $\theta $ such that
\begin{equation}
  \label{bar}
XF\Phi _{m}^{(\theta )}=e^{2\pi i\theta _{1}}F\Phi _{m}^{(\theta )}
\end{equation}
and
\begin{equation}
  \label{ugh2}
YF\Phi _{m}^{(\theta )}=e^{2\pi i\theta _{2}}F\Phi _{m}^{(\theta )}
\textnormal{.}
\end{equation}

We shall use $Y^{1/2}=e^{i\widehat{p}/2\hbar }$. The first thing to do is
get the commutation relations straight:
\begin{eqnarray*}
Y^{1/2}L &=&RY^{1/2},\,Y^{1/2}X=\left( -1\right) ^{N}XY^{1/2}\textnormal{,} \\
X^{-1}E_{p} &=&O_{p}X^{-1},\,LR=E_{p}O_{p}=0.
\end{eqnarray*}
Thus, we find
\begin{eqnarray*}
XF\Phi _{m}^{(\theta )}&
=&XS(L+X^{-1}R)(E_{p}+Y^{-1/2}O_{p})\Phi _{m}^{(\theta )}  \nonumber \\
&=&S(L+X^{-1}R)(E_{p}+Y^{-1/2}O_{p})X^{2}\Phi _{m}^{(\theta )}
\\
&=&e^{2\pi i\left( 2\theta _{1}\right) }F\Phi _{m}^{(\theta )}
\end{eqnarray*}
Comparing with eqn. \ref{bar} we see that $\theta _{1}=0$.
Also
\begin{eqnarray*}
&&YS(L+X^{-1}R)(E_{p}+Y^{-1/2}O_{p})\Phi _{m}^{(\theta )}   \\
&=&S(R+\left( -1\right) ^{N}X^{-1}L)(E_{p}+Y^{-1/2}O_{p})
Y^{1/2}\Phi_{m}^{(\theta )} \\
&=&S(\left( -1\right) ^{N}XR+L)
(O_{p}+\left( -1\right) ^{N}Y^{-1/2}E_{p})
Y^{1/2}\Phi_{m}^{(\theta )},
\end{eqnarray*}
where we have used $\theta_1 = 0$ in the last step. 
 Substituting this
into eqn. \ref{ugh2}, we find
\begin{eqnarray*}
S(\left( -1\right) ^{N}XR+L)
(e^{2\pi i \theta _{2}}Y^{-1/2}O_{p}+\left( -1\right) ^{N}E_{p})
\Phi_{m}^{(\theta )}  \\
=e^{2\pi i\theta _{2}}S(L+X^{-1}R)(E_{p}+Y^{-1/2}O_{p})\Phi _{m}^{(\theta )}
.
\end{eqnarray*}
That is,
\begin{eqnarray*}
0 &=&\left( YF-e^{2\pi i\theta _{2}}F\right) \Phi _{m}^{(0,\theta _{2})} \\
&=&\left( \left( -1\right) ^{N}-e^{2\pi i\theta _{2}}\right) 
SLE_{p}\Phi _{m}^{(0,\theta _{2})}
\\
&&+\left( 1-e^{2\pi i\theta _{2}}\right) SX^{-1}RE_{p}
\Phi_{m}^{(0,\theta _{2})} \\
&&+e^{2\pi i\theta _{2}}\left( \left( -1\right) ^{N}-1\right) 
SX^{-1}RY^{-1/2}O_{p}
\Phi_{m}^{(0,\theta _{2})}.
\end{eqnarray*}

Applying the operator $LS^{-1}$ on the left gives the condition
$\left( -1\right)
^{N}=e^{2\pi i\theta _{2}}$ so that $\theta _{2}=0$ if $N$ is even and $%
\theta _{2}=1/2$ if $N$ is odd. For the even case, it follows that
$\theta =\left( 0,0\right) $ is the only invariant point. For the odd
case, we are left with
\begin{eqnarray*}
&&\left( YF-e^{2\pi i\theta _{2}}F\right) \Phi _{m}^{(0,\theta _{2})} \\
&=&2SX^{-1}R\left( E_{p}+Y^{-1/2}O_{p}\right) \Phi _{m}^{(0,1/2)}
\textnormal{,}
\end{eqnarray*}
and this operator will never vanish everywhere.

\paragraph*{Discussion}

What this means is that for the propagator presented in ref. \cite{RS}, it
is impossible to construct an $N$-dimensional subspace which is
invariant under $U$, $V$ and $F$ except for the case
of $N$ even and periodic boundary conditions.

We were led to look for non-periodic invariant points by the work of Saraceno
(ref. \cite {S}).  He has proposed an ``anti-periodic'' quantization
which differs from both eqn. \ref{OURS} and the Balazs-Voros
quantization by a correction of order $\hbar $.  That work defines $U$
and $V$ similarly to the definitions in \cite{RS} and demands that
$U^N = V^N = -1$.  In the terminology we use here, this would be at
$\theta_1 = \theta_2 = 1/2$.  The preceding calculation indicates that
this point is not an invariant point of the propagator given by
eqn. \ref{propagator}.

In fact, the general impossibility of finding this point invariant is
evident by considering the momentum basis of the $\theta = (1/2,1/2)$
eigenspace.  
\begin{eqnarray*}
  \tilde\Phi_n^{(\theta_1,\theta_2)} = e^{-2\pi i n \theta_1/N}
  \sum_k e^{-2\pi i \theta_1 k}
  \left| \frac{\theta _{2}+n}{N} +k\right\rangle _{p}
\end{eqnarray*}
Under the bakers map, for $n = 0$, the states centered at 
$p = {\theta _{2}}/{N} +k$ would go to states centered at 
$p = {\theta _{2}}/({2N}) +{k}/{2}$ which are not in our
subspace for $\theta_2 \ne 0$.  Thus, even under the classical baker's
map,  the region of phase space corresponding to states with
antiperiodic boundary conditions does not map into itself.

While it might be possible to change the quantum propagator
(eqn. \ref{propagator}) in order
to make $\theta = (1/2,1/2)$ be a fixed point, it is hard to imagine
what  could be done for general $\hbar$ which would have just the right
effect for $h=1/N$.  We are thus led to believe that the quantization
in ref. \cite{S}, while drawing inspiration from anti-periodic states
does not have a direct analogue within the more explicit quantization
approach of ref. \cite{RS}.

%\bibliography{local}

\end{document}